\documentclass[pdflatex,sn-nature]{sn-jnl}

\usepackage{graphicx}%
\usepackage{multirow}%
\usepackage{amsmath,amssymb,amsfonts}%
\usepackage{amsthm}%
\usepackage{mathrsfs}%
\usepackage[title]{appendix}%
\usepackage{xcolor}%
\usepackage{textcomp}%
\usepackage{manyfoot}%
\usepackage{booktabs}%
\usepackage{algorithm}%
\usepackage{algorithmicx}%
\usepackage{algpseudocode}%
\usepackage{listings}%
\usepackage{array}%
\usepackage{url}%

\usepackage{cleveref}
\usepackage[]{hyperref}

\newcommand{\ie}{i.e.~ }

\newcommand{\der}[2]{\frac{\mathrm{d} #1}{\mathrm{d}#2}}

\newcommand{\figref}[1]{Fig.~\ref{#1}}
\newcommand{\eqnref}[1]{Eq.~\ref{#1}}
\newcommand{\tabref}[1]{Tab.~\ref{#1}}

\newcommand{\ev}[1]{\langle {#1} \rangle}

\begin{document}

\title[Dispersal diversity buffers species vulnerability to local extinction]{Dispersal diversity buffers species vulnerability to local extinction}

\author[1,2]{\fnm{Davide} \sur{Bernardi}}
\author[3]{\fnm{Giorgio} \sur{Nicoletti}}
\author[4]{\fnm{Prajwal} \sur{Padmanabha}}
\author[1]{\fnm{Samir} \sur{Suweis}}
\author[1,2]{\fnm{Sandro} \sur{Azaele}}
\author[5,6]{\fnm{Simon A.} \sur{Levin}}
\author[7,8]{\fnm{Andrea} \sur{Rinaldo}}
\author[1,2]{\fnm{Amos} \sur{Maritan}}

\affil[1]{\orgdiv{Laboratory of Interdisciplinary Physics, Department of Physics and Astronomy}, \orgname{University of Padova}, \orgaddress{\city{Padova}, \postcode{35131}, \country{Italy}}}
\affil[2]{\orgname{National Biodiversity Future Center}, \orgaddress{\city{Palermo}, \postcode{10587},  \country{Italy}}}
\affil[3]{\orgdiv{Quantitative Life Sciences}, \orgname{The Abdus Salam International Centre for Theoretical Physics}, \orgaddress{\city{Trieste}, \postcode{34151}, \country{Italy}}}
\affil[4]{\orgdiv{Department of Fundamental Microbiology}, \orgname{ University of Lausanne}, \orgaddress{\city{Lausanne}, \postcode{1015}, \country{Switzerland}}}
\affil[5]{\orgdiv{Department of Ecology and Evolutionary Biology}, \orgname{Princeton University}, \orgaddress{ \city{Princeton}, \postcode{08544}, \state{NJ}, \country{USA}}}
\affil[6]{\orgdiv{High Meadows Environmental Institute}, \orgname{Princeton University}, \orgaddress{\city{Princeton}, \postcode{08544}, \state{NJ}, \country{USA}}}
\affil[7]{\orgdiv{CMCC Foundation}, \orgname{Euro Mediterranean Center on Climate Change}, \orgaddress{\city{Lecce}, \postcode{73100}, \country{Italy}}}
\affil[8]{\orgdiv{Faculty ENAC}, \orgname{\'Ecole Polytechnique F\'ed\'erale de Lausanne}, \orgaddress{\city{Lausanne}, \postcode{1015}, \country{Switzerland}}}

\abstract{Predicting species persistence within ecological communities is a fundamental challenge for both empirical and theoretical ecology. Existing methods span from  mechanistic models, whose parameters are difficult to estimate from data, to statistical tools whose context-specific parameters are less interpretable.
Here, we present a general framework, grounded in the statistical physics of complex systems, that integrates the key processes governing species survival into a single measurable quantity: the competitive balance. This metric quantifies a focal species' vulnerability to competitive exclusion beyond what is captured by its abundance alone by incorporating the diversity of dispersal strategies and the structure of interspecific interactions within the community. Crucially, it can be inferred from spatial abundance data, thus circumventing the need to estimate species traits or dispersal parameters.
Our results reveal that greater heterogeneity in dispersal strategies reduces vulnerability to competitive exclusion for a given abundance. Although we validate the framework using tropical and temperate forest data, it can be applied to a range of different ecosystems, providing a systemic and interpretable tool for assessing a context-dependent species vulnerability that accounts for its interactions with the entire community.}

\maketitle

\noindent
Human impacts on biodiversity have prompted concerns about a potential mass extinction \cite{mass_extinction_nature_Barnosky_2011}. Yet, despite signals such as population declines and range contractions, the magnitude and the trajectory of ongoing extinctions remain debated \cite{CowBou2022}. Resolving these uncertainties requires theoretical frameworks that link local population dynamics to spatial community structure across scales \cite{levin2000multiple}, and that can be ultimately validated with empirical data \cite{tilman2017future}.

Ecological theories have made substantial progress on different aspects of this problem. Coexistence theory details broad mechanisms by which species persist together \cite{chesson2000mechanisms,Saavedra_mutualistic_bascompte2011_nature,Levine_allesina_review_2017,Bimler2025}, with recent work extending these ideas to empirical measurement \cite{Ellner2019}, demographic uncertainty \cite{Bowler2022}, and community-level characterizations of the feasibility regions \cite{Grilli2017,Spaak2020}. Metapopulation \cite{hanski1998} and metacommunity \cite{leibold2004metacommunity} frameworks have shown how spatial structure and dispersal shape diversity \cite{Grainger2016}, while broader theoretical syntheses have connected biodiversity to ecosystem stability \cite{loreau2013biodiversity}, life-history variation \cite{Jops2023oDwyer}, and macroecological patterns \cite{Harte_2014}. Yet, these advances have largely proceeded in parallel: analytical tractability for species-rich communities has been achieved mainly in non-spatial settings \cite{allesina2012stability,Grilli2017}, while spatially explicit models remain difficult to scale to many interacting species.

Spatial heterogeneity shapes coexistence in multiple ways: through stabilizing mechanisms such as the storage effect and dispersal limitation \cite{Levin1992,spatial_coex_Hart_levine_2017},   by giving rise to interspecific spatial associations \cite{Keil2021}, and by modulating coexistence–area relationships through the configuration of species interactions \cite{GarciaCallejas2021}. Empirical tools for quantifying these patterns include spatial point pattern analysis \cite{Wiegand2016} and joint species distribution models \cite{Harris2015}.  Recent work has begun to integrate spatial patterns with interaction-based descriptions of communities \cite{Wiegand_2021,Kalyuzhny_2023,Wiegand_2025}. Yet, integrating spatial structure into mechanistic models of interacting communities remains challenging. Analytical results typically become intractable, while numerical simulations depend sensitively on model assumptions and parameterization, and offer limited scalability \cite{Pilowsky2022review}. 

On the empirical side, data-driven methods (such as species distribution models \cite{Elith2009SDM}, occupancy scaling methods \cite{kunin1998science,Wilson_2004Kunin}, Population Viability Analysis \cite{Shaffer1981viability}, or Matrix Population Models \cite{Vindenes2021MPM}) combine statistical inference with some mechanistic elements, but often focus on single-species dynamics and typically require detailed demographic and environmental data \cite{Chaudhary2020PVA,Simmonds2024MPM}. Hence, there is a need for frameworks that combine spatial structure with multispecies interactions in a way that is analytically tractable and compatible with the coarse-grained data available for most species.

Here, we introduce an analytical framework to quantify each species' proximity to its competitive exclusion boundary in  metacommunities. Our approach employs a parsimonious ecological model explicitly incorporating basic processes taking place in a heterogeneous spatial environment. Leveraging analytical methods from the physics of disordered systems, we derive a set of ecologically interpretable metrics that quantify how species exploit locally available space and how they are affected by community-mediated interactions. These metrics define species vulnerability as their proximity to the competitive-exclusion threshold, explicitly accounting for the ecological context in which they are embedded. Importantly, our approach does not require prior knowledge of species-specific traits, demographic parameters, or interaction coefficients, and can be applied directly to empirical data. This is especially valuable in studying species-rich communities, where detailed species-level information is often unavailable and might be strongly context-dependent. A summary of the main metrics and definitions used throughout the study is provided in \cref{tab:definitions}.

Backed by comparison to forest datasets, we exploit our framework to show that species vulnerability cannot be fully predicted by abundance alone.
Rather, we explicitly quantify how the entanglement among dispersal strategies, spatial distribution patterns, and total population size determines species’ proximity to exclusion from a metacommunity, capturing an ecosystem-level perspective on long-term persistence  that moves beyond single-species metrics and risk assessments based solely on demographic fluctuations.
Further, although our metrics are derived analytically, they remain interpretable without relying on model specifics, suggesting broader ecological relevance.

\subsection*{Statistical physics-based theory of spatially-explicit ecosystems}
We study a general metacommunity model describing the temporal dynamics of species abundances in a habitat composed of $N$ colonizable sites, each providing a finite amount of available space or a limiting resource. These sites may correspond to distinct patches, although our framework does not strictly require explicit physical boundaries. The model incorporates basic processes such as death, reproduction, dispersal, colonization, and competition for space. Their combined effect is captured by an interaction-dispersal kernel matrix, $\boldsymbol{K}$, whose elements represent a coarse-grained description of how species interact and spread across patches. However, they do not explicitly separate local competitive and facilitative interactions, but capture instead their net effect on dispersal efficacy.

\renewcommand{\arraystretch}{1.6}  
\begin{table}
\centering
\footnotesize
\begin{tabular}{|m{0.64\textwidth}|m{0.33\textwidth}|}
\hline
Description & {Symbol / mathematical definition}\\
\hline
\hline
\textbf{Space-use fraction}. Fraction of the total effective space available in site (or subplot) $i$ occupied by species $\alpha$. For forest data, the effective space is estimated from the allometric relation between trunk diameter and crown area \cite{Simini2010pnas,MartinezCano2019}. If diameter data are not available, it can be approximated using stem counts.
& $p_{\alpha i}= \sum_{j} d_{i\alpha,j}^{4/3}/M$\newline \newline $d_{i\alpha,j}$ is the trunk diameter of the $j$th individual belonging to species $\alpha$ at site $i$\newline $M$ is the total effective space per site. It is estimated from the data from a global fitting procedure (see Methods)  \\
\hline
\textbf{Total fraction of occupied space.} Average space-use fraction, or total fraction of effective available space occupied by species $\alpha$ (across the entire sampled area).
& $\ev{p_{\alpha}}$, where $\ev{.}$ indicates average over subplots (sites) \\
\hline
\textbf{Vacancy-Adjusted Abundance (VAA).} Space-use of species $\alpha$ in subplot $i$, normalized by the locally available (unoccupied) space. By accounting for local space saturation, the VAA rescales relative space occupation to reflect the effective colonization effort.  As a result, identical relative abundances yield higher VAA values in space-limited sites and lower values in less saturated sites. The dynamical mean field calculation provides an analytical expression for the VAA distribution that can be fitted to spatial census data.
& $x_{\alpha i} = \frac{p_{\alpha i}}{1 - \sum_\beta p_{\beta i}}$ \\
\hline
\textbf{Interaction-dispersal kernel.} Species-specific matrix describing interaction and dispersal processes of species $\alpha$. The kernel matrix $\mathbf{K}_\alpha$ bundles into a single quantity local birth and growth rates, colonization-mediated dispersal, and local interactions of species $\alpha$ with other species, such as competition and facilitation (in the form of effective competitive suppression). Its spatial distribution is assumed to be a species-specific Gamma distribution. It is not directly measurable, but its parameters are inferred from the data by fitting the analytical theory for the VAA to the empirical VAA distribution. 
& $\mathbf{K}_\alpha$\\
\hline
\textbf{Dispersal heterogeneity.} Spatial variability of the dispersal strategy. The dispersal heterogeneity of species $\alpha$ is defined as the inverse shape parameter of the VAA distribution, $\delta_\alpha^{-1}$,  which is equal to the squared coefficient of variation of the VAA distribution. High $\delta_\alpha^{-1}$ reflects patchy, spatially aggregated abundance; low $\delta_\alpha^{-1}$ reflects a more spatially homogeneous distribution. 
& $\delta^{-1}_\alpha$ \\
\hline
\textbf{Effective dispersal strength.} Scale (or inverse rate) parameter of the VAA distribution. The theory links it to the average of the $K_\alpha$ distribution, adjusted by the average abundance and local death rate (see Methods). It is related to the effective dispersal strength of the species, relative to the community level.
& $\bar{\beta}^{-1}_\alpha $ \\
\hline
\textbf{Competitive balance.} Ratio between community-wide factors that increase and decrease abundance. $B_\alpha > 1$ is necessary for persistence in isolation; persistence of species $\alpha$ in the community requires $B_\alpha$ to be not too far below the equal-balance condition. The competitive balance is inferred indirectly from data via a self-consistent integral equation coupling all species' VAA parameters. This equation does not additively sum the competitive balances of all species, but it determines the metacommunity composition through an implicit nonlinear equation.  
& 
$$\begin{aligned}
B_{\alpha} \!=\! \left(\!\int_{0}^{\infty}\!\! d\lambda  \frac{e^{-\lambda
			\!- \!\sum_{\gamma}\! \delta_{\gamma} \ln \left(1 + {\lambda}/({{\bar{\beta}}_{\gamma} )}\right)}}{ \left.1 + {\lambda}/{{\bar{\beta}}_{\alpha} }\right.}	\!\right)^{\!\!\!-1}
\end{aligned}$$
\\
\hline
\textbf{Species vulnerability.} Rescaled distance to each species' persistence threshold. The critical value is zero. Species with negative vulnerability persist in the long term; thus, a more negative vulnerability (i.e., larger $|W_\alpha|$) indicates lower susceptibility to competitive exclusion. Species vulnerability is computed from the competitive balances and the kernel parameters of \emph{all} species, so it depends on the ecological context. The same species can have a different vulnerability in different communities, depending on their composition.
& $W_\alpha= \frac{1}{|\Delta_\alpha|}\bigg(\big(\left[B_\alpha\right]_\alpha- B_{\alpha}\big)  -\Delta_\alpha \bigg)$  \newline \newline
$\left[.\right]_\alpha$ indicates average over species; \newline \newline$\Delta_\alpha$ is the rescaled distance of the community average competitive balance to the persistence threshold of species $\alpha$. See Methods, \cref{eq:species-vulnerability-1} for its explicit expression.  \\
\hline
\end{tabular}
\caption{\textbf{Metrics and definitions introduced in this study. }}
\label{tab:definitions}
\end{table}
\renewcommand{\arraystretch}{1.}  

The mathematical structure of the model aligns with the metacommunity framework \cite{hanski1998}, widely used to study species survival in fragmented landscapes \cite{HanOva2000, mouquet2002coexistence, PadNic2024}. However, measuring interaction kernels or the corresponding dispersal networks in real ecosystems remains challenging. In light of this experimental inaccessibility, we posit that the intrinsic variability in habitats, species traits, and interspecific interactions in large-scale natural ecosystems will lead to entries in the kernel that are effectively random. We model this randomness by drawing kernel elements from a Gamma distribution (Methods). The parameters of this distribution are species-specific, meaning that each species has its own parameter pair, that quantitatively capture diverse colonization strategies. These parameters are not prescribed a priori, but inferred directly from observed spatial abundance data.

\begin{figure}
		\centering
		\includegraphics[width=\textwidth]{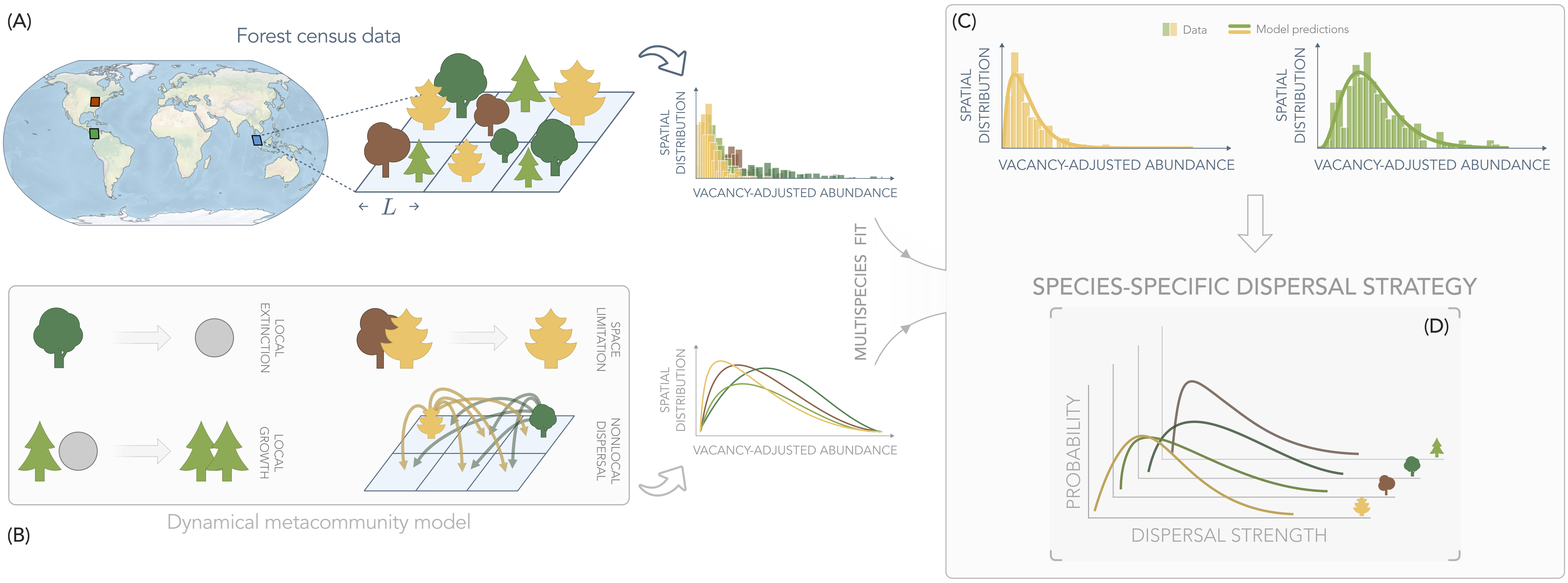}
		\caption{\footnotesize\textbf{Fitting theory-predicted vacancy-adjusted abundance distributions (VAAD) to data enables extraction of species-specific interaction-dispersal strategies.}		
\textbf{(A)}: Spatial tree census data are aggregated into square subplots of size $L$ and converted into species-specific local space-use fractions $p_{\alpha i}$, where $\alpha$ denotes species and $i$ spatial location. These are normalized by the locally unoccupied space --estimated during fitting-- to obtain the empirical VAAD, which rescales relative abundance to account for local space saturation and capture colonization effort (see \cref{tab:definitions}).
\textbf{(B)}: Using methods from statistical physics applied to a parsimonious metacommunity model with growth, death, competition for space, and dispersal, we derive an analytical prediction for the VAAD (Methods). 
\textbf{(C)}: The parameters of the theoretical VAAD are linked to those of the interaction–dispersal kernel through an implicit integral equation. Fitting the theory to data therefore enables inference of this kernel, which encodes species-specific spatial strategies (D). Hence, this inference does not require direct measurements of vital rates, demographic rates, or interaction coefficients.}
\label{fig:model-new}
\end{figure}

The link between model and data is provided by a metric that emerges from the analytical solution of the model using generalized dynamical mean field theory (DMFT) \cite{AzaMar2024}: the \emph{vacancy-adjusted abundance} (VAA). 
The VAA of species $\alpha$ at a given site is defined as the fraction of space it occupies divided by the locally available (unoccupied) space. This normalization accounts for local space saturation,  making the VAA dependent on how empty a local site is relative to its capacity. Therefore, a species with the same relative space-use in different subplots contributes differently to the VAA depending on how much each patch is occupied: a subplot with very low free space amplifies the VAA for species occupying this spot, and it reduces it in subplots that have larger free space.
Although the VAA arises naturally from the DMFT calculation (Methods), it has a clear ecological interpretation: by upweighting space-limited patches, it rescales local relative space occupation to reflect the effective colonization effort.

For each species, the parameters of the VAA distribution quantify its dispersal heterogeneity ($\delta_\alpha^{-1}$ in \cref{tab:definitions}) and an effective dispersal strength ($\bar\beta_\alpha^{-1}$ in \cref{tab:definitions}), which depends on the surrounding community. Fitting empirical VAA distributions therefore enables inference of species-specific dispersal strategies without requiring direct measurements of dispersal or interaction parameters.
Because the parameters of the theoretical VAAD are linked to those of the interaction-dispersal kernel through an implicit integral equation, fitting the theory to data enables inference of this kernel, which encodes species-specific spatial strategies. In this way, the VAA distribution provides a direct link between observable space-use patterns and the underlying interaction-colonization structure, while remaining a model-independent quantity that can be estimated from spatial abundance data. The overall conceptual pipeline is illustrated in \figref{fig:model-new}.

\subsection*{Theory predicts species-specific VAA distributions in forest plots}
We validate our theory on three forest datasets: the Barro Colorado Island (BCI) tropical forest plot in Panama~\cite{BCI_data}, the Pasoh Forest Reserve tropical forest plot in Malaysia~\cite{manokaran2004pasoh}, and the Michigan Big Woods temperate forest plot in the USA~\cite{michigan_data}. Each dataset provides a spatially complete census of all trees and thus offers an ideal testbed for our framework, which models spatially explicit abundances of species competing for shared space.

We computed the VAA empirical distributions from data, as described in Methods. Briefly, we partition each plot into square subplots of side length $L=50$~m, a spatial aggregation scale that balances sufficient sampling within subplots with the need to retain spatial resolution (SI Sec.~1 and Supplementary Fig.~S2, for details of optimal scale selection); we then estimate the space used by each species in each subplot by accounting for each tree's size, following the established allometric scaling relation between trunk and crown size \cite{Simini2010pnas,MartinezCano2019}; next, we infer the per-subplot effective total space capacity using a global fitting procedure. Finally, we construct the empirical VAA histogram as per its definition (see \cref{tab:definitions}). 

\begin{figure}
		\includegraphics[width=1.05\textwidth]{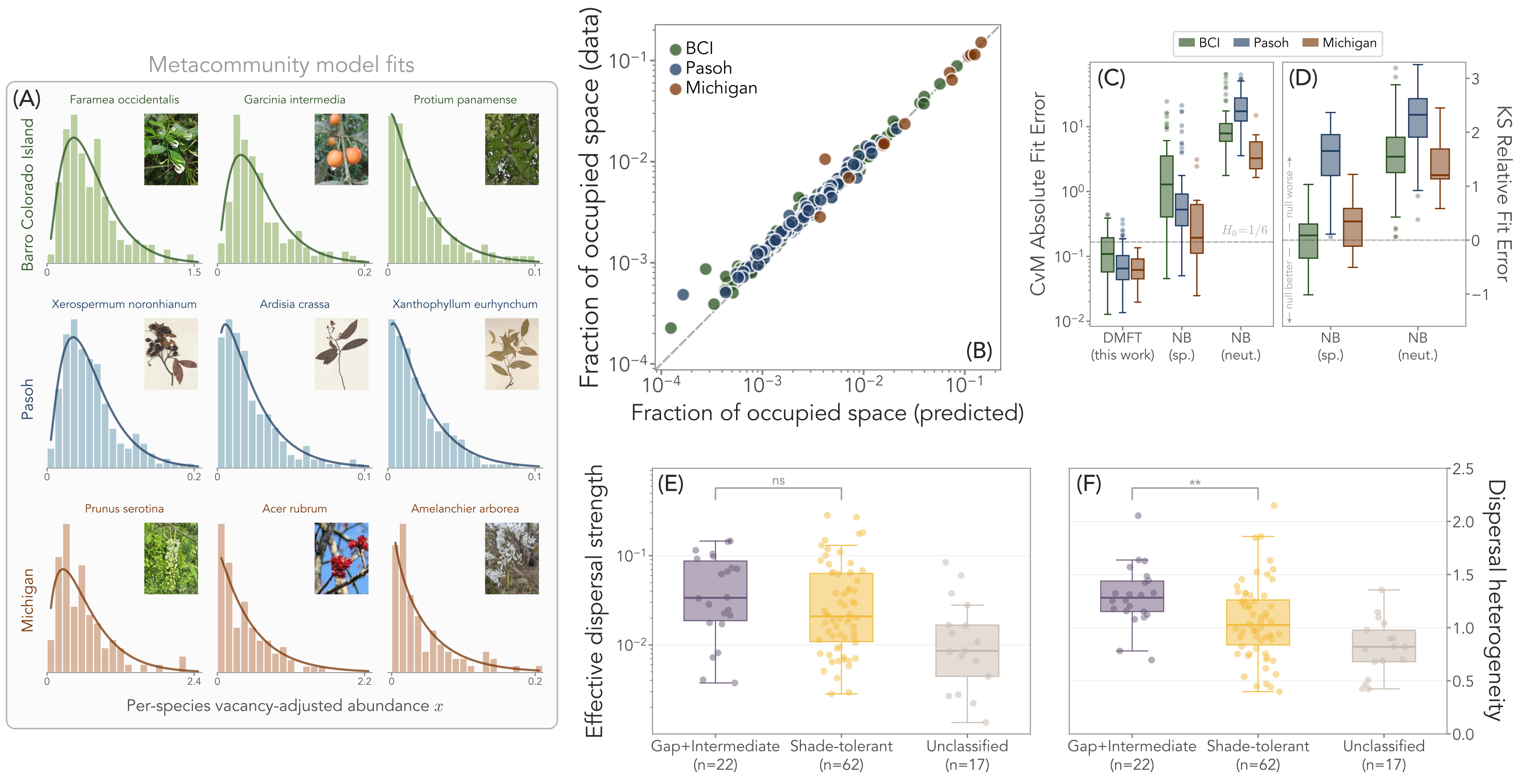}
		\caption{\footnotesize\textbf{Theory accurately predicts species-specific vacancy-adjusted abundance (VAA) distributions from forest tree census data. VAA parameters reflect species dispersal and life-history strategies.}
			\textbf{(A)}: Empirical VAA distributions (histograms) with theory (lines) and for representative species from the Barro Colorado Island (BCI), Pasoh, and Michigan forest plots (full report in SI); species photo credits in Supplementary Table~S3.
			\textbf{(B)}: Out-of-sample theoretical prediction for mean fraction of space-use by species $\alpha$, $\langle p_\alpha\rangle$, vs.\ observed $\langle p_\alpha\rangle$. Points closely follow the identity line across all three datasets (for BCI, Pasoh, and Michigan: $R^2=0.97, 0.93, 0.95$; $n = 100$, $100$, $12$ number of species, respectively). \textbf{(C), (D)}: Comparison with non-interacting overdispersed null models based on negative binomial fitting of species counts (see Methods and SI, sec.~6). Two null model variants are considered: a neutral variant with the same fit parameters for all species and a variant with species-specific fit parameters.  Our model not only strongly outperforms the neutral model, but also performs better than species-specific model where the number of parameters are matched. The absolute fit error (C) is based on a CvM proxy (dashed line indicates expectation for a correctly specified model), and the relative fit error (D) measures the log ratio of the Kolmogorov-Smirnov statistic (positive ratio means that the null model is worse; see Methods and SI, sec.~6 for details). In (C) and (D), each box summarises the per-species distribution of the fit-error metric over the $n = 100$, $100$ and $12$ most abundant analysed species (BCI, Pasoh and Michigan, respectively; the same species shown in panel B), coloured by dataset (colour key in panel).
			\textbf{(E)}: Effective dispersal strength ($\bar \beta^{-1}_\alpha$) and \textbf{(F)}: dispersal heterogeneity ($\delta_\alpha^{-1}$) by ecological guild for BCI data ($n = 22$ Gap+Intermediate, $62$ Shade-tolerant and $17$ Unclassified species; guild assignments from ref.~\cite{Comita2007}). Guilds were compared with two-sided Mann--Whitney $U$ tests, Holm-corrected for multiple comparisons; Unclassified species are displayed but excluded from testing. Gap and intermediate species have significantly higher dispersal heterogeneity $\delta_\alpha^{-1}$ than shade-tolerant species (F: $U = 969$, $p = 0.004$, \textbf{**}), indicating that shade-tolerant species display more spatially homogeneous abundance distributions, whereas effective dispersal strength $\bar\beta_\alpha^{-1}$ does not differ significantly between the two groups (E: $U = 815$, $p = 0.18$, ns). Significance markers: ns $p > 0.05$, \textbf{*} $p \le 0.05$, \textbf{**} $p \le 0.01$. Box plots (C--F): the centre line is the median, the box spans the interquartile range (IQR; 25th--75th percentiles), and whiskers extend to the most extreme value within $1.5\times$IQR of the box; in (C) and (D) points beyond the whiskers are drawn as faded dots (outliers), while in (E) and (F) all individual species are overlaid as jittered dots.}
		\label{fig:VAAD-fits}
\end{figure}

\figref{fig:VAAD-fits}A compares the theoretical VAA predictions, fitted to the empirical distributions, for three representative species per dataset. We observe excellent agreement, with the model accurately capturing the diversity of distributional shapes across species. These differences are naturally accounted for by species-specific dispersal strategies (Methods). Goodness-of-fit analyses for all analyzed species, along with explicit visualizations for 212 species (Supplementary Figs.~S3--S5), are reported in SI Sec.~1.

Next, we performed a series of model robustness checks. First, because the net effect of space limitation is competitive, we tested whether explicitly incorporating facilitation \cite{McIntire2014,Bimler2025} -- beyond the competition buffering implicitly encoded in the inferred parameters of $\mathbf{K}$ -- could improve model performance. This extension resulted in only marginal changes in model fit across the three forest datasets (SI, sec. 5). This suggests that, at the spatial and temporal scales considered here, the dominant effects of facilitation may already be effectively captured by the inferred kernel parameters.

Then, we investigated whether our model generalizes to unseen data. Specifically, we used the parameters inferred on half of the forest plot (with different random partitions) to predict the mean fraction of space occupied by each species in the remaining half. This out-of-sample validation of our theory resulted in a remarkable agreement for all three datasets (\figref{fig:VAAD-fits}B), showing that our analytical predictions are well-aligned with empirical observations.

As a final check, we compared our framework against ecological null models in which species are independent and overdispersed, leading naturally to a Negative Binomial description of abundances \cite{Linden2011}. We considered two variants (Methods): a neutral model parameterized at the community level, closely related to neutral theories of biodiversity that reproduce several large-scale ecological patterns in species-rich systems \cite{Volkov2003}, and a species-specific model fitted independently for each species, matching the total number of parameters of our framework.
We evaluated model performance by fitting the VAA distributions of all species across the three datasets (see Methods). For each dataset, the absolute goodness-of-fit of our model is consistent with the reference level expected for a correctly specified model (\figref{fig:VAAD-fits}C, dashed line). In contrast, the species-specific model provides an adequate fit only for the Michigan dataset, while the neutral model exhibits substantially larger errors across all datasets, suggesting model misspecification.
In a relative goodness-of-fit comparison, our framework consistently outperforms the neutral model across all datasets (positive KS log-ratio in \figref{fig:VAAD-fits}D). Relative to the species-specific model, it yields a marked improvement in Pasoh and Michigan, and a comparable, slightly better performance in BCI. Full details are provided in the SI (Sec.~6), where an intermediate variant with partially shared parameters is also analyzed. Additional analyses of data requirements (SI, Sec.~7) show that reliable parameter estimation is achievable even for smaller datasets and provide lower-bound estimates on data requirements.

Crucially, besides achieving a better fit to the VAA distributions, our model also yields two species-specific parameters that carry clear ecological information: dispersal heterogeneity and effective dispersal strength (\cref{tab:definitions}). Based on the meaning of the VAA, we hypothesize that VAA heterogeneity also reflects differences in life-history strategies. To test this, we compared parameter values across functional guilds in the BCI dataset (shade-tolerant, gap-exploiting, and intermediate; classification from ref.~\cite{Comita2007}), focusing on a grouped comparison of gap-exploiting and intermediate species versus shade-tolerant species.
We found no statistically significant difference between guilds in the effective dispersal strength (\figref{fig:VAAD-fits}E; see caption for details on the statistical test). In contrast, gap-exploiting and intermediate species exhibit significantly larger dispersal heterogeneity than shade-tolerant species (\figref{fig:VAAD-fits}F). This difference between gap exploiters and shade-tolerant species remains significant when treating all three guilds separately (SI, Sec.~4). These results are consistent with the role of canopy gap dynamics in promoting episodic local dominance of light-demanding species \cite{BrokawBusing2000,Kupers2019}, and with recent work highlighting the role of life-history strategies in shaping species responses to environmental variability and enabling coexistence \cite{Jops2023oDwyer,jops2025oDwyer}.

\subsection*{Competitive balance as the keystone of species coexistence}
Our theoretical predictions go beyond capturing the shape of VAA distributions and connecting their parameters to species' dispersal strategies; they also reveal how these parameters relate to the long-term persistence of species within the community. In classic metapopulation models that focus on a single species and do not model others explicitly, persistence occurs when the local extinction rate is lower than the so-called metapopulation capacity, typically quantified by the largest eigenvalue of the kernel matrix \cite{HanOva2000,NicPad2023}. However, when multiple competing species are considered, the metapopulation capacity of each species provides a necessary but not sufficient condition for persistence \cite{PadNic2024}. In our case, we show that species persistence within an interacting community is determined by how a specific combination of single-species traits, which we term \emph{competitive balance}, compares to the competitive balance of all other species (Methods).

Our calculations provide an explicit integral equation that connects the competitive balance of each species to the VAA distribution parameters of all species within the community (see \tabref{tab:definitions}). The structure of this equation has two important consequences: first, competitive balances are not simply additive, but coupled non-linearly through a shared term that makes community-level coexistence emerge as a collective property rather than a sum of individual contributions; second, it enables the inference of all competitive balances from data. Intuitively, competitive balance reflects how community-wide processes influence each species-specific ratio between factors that increase abundance and those that decrease it. A species can persist if its competitive balance is both greater than one, indicating that it could at least sustain itself in the absence of competition, and not too far below an \emph{equal balance condition}, a hypothetical scenario in which all species share identical balance values.

\begin{figure}
	\centering
	\includegraphics[width=\textwidth]{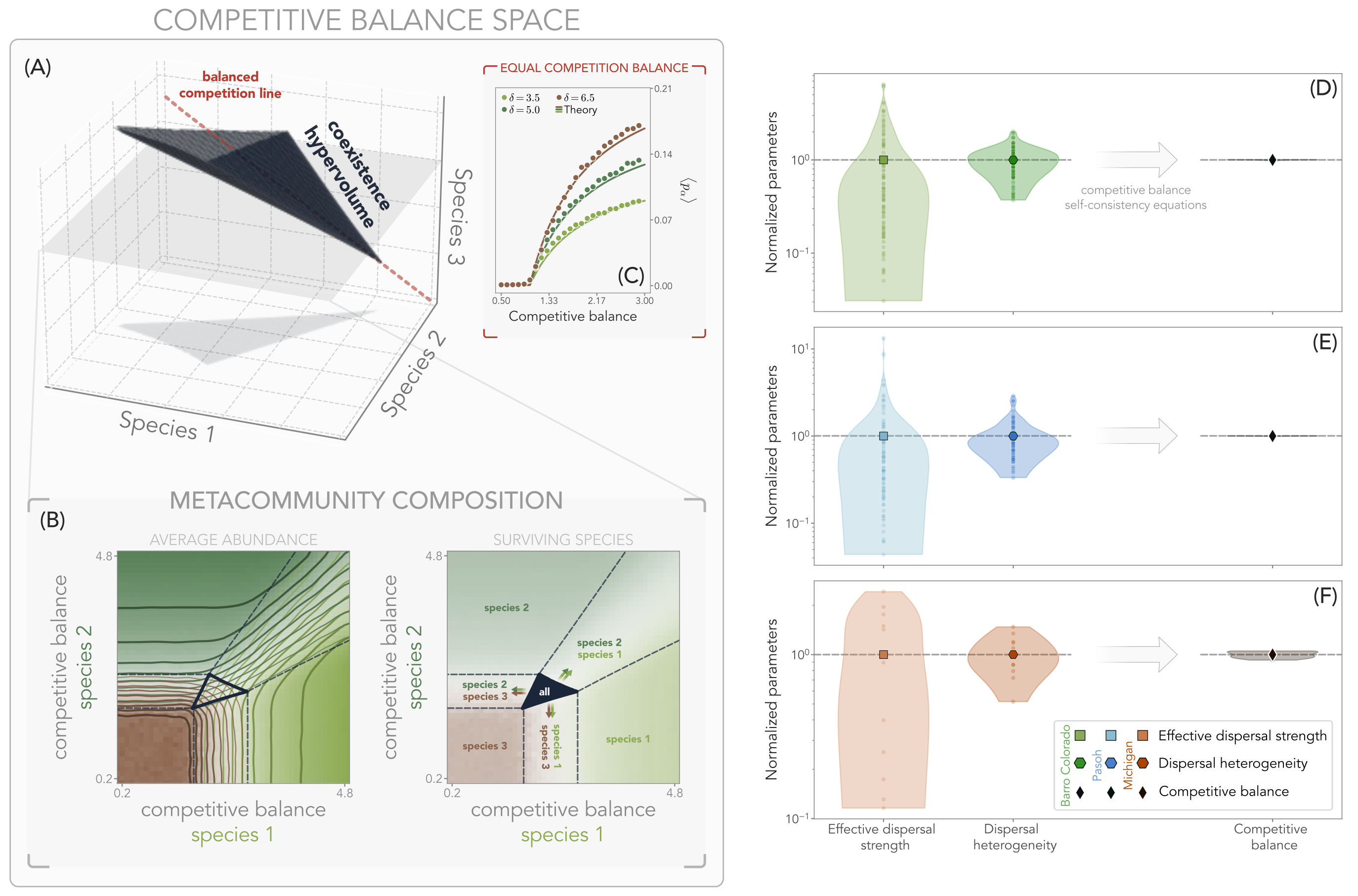}
	\caption{\footnotesize\textbf{Theory reveals emergent colonization-extinction trade-off and identifies parameters governing survival and coexistence in species-rich communities.}
		(\textbf{A}) In the space defined by competitive balance values, coexistence of $S$ species is possible only within a region where species' balances are sufficiently similar. Here $S=3$ for visualization.  (\textbf{B}):  A two-dimensional slice of this space shows species abundances (color intensity and contour lines) from numerical simulations as a function of the competitive balance of species 1 and 2, with species 3 held fixed. Theoretical predictions (critical boundaries) partition this space into distinct coexistence regimes. (\textbf{C}):   Simulations (circles) and theory (lines) show that species with identical competitive balances can have different abundances depending on community context (here the species have different dispersal heterogeneity, see \tabref{tab:definitions}). (\textbf{D}): Species-specific parameters -- effective dispersal strength and dispersal heterogeneity (from VAA fits) -- and competitive balances (computed from fitting parameters using the theory) are plotted relative to their means for BCI data. Each dot represents a species, with distributions visualized via kernel density estimates. While dispersal-related parameters vary widely, competitive balances are tightly clustered around their mean, consistent with theoretical expectations. (\textbf{E}): Same as (D) but for Pasoh data. (\textbf{F}): Same as (D, E) for Michigan forest data.}
	\label{fig:balance-and-violins}
\end{figure}

This concept is illustrated in \figref{fig:balance-and-violins}A using the example of three interacting species. In this representation, an ecological community corresponds to a point in a coordinate space defined by the competitive balance of each species. Our theory predicts the hypervolume in the competitive balance space where all species coexist, as well as the regions where all possible subsets of species coexist. Moreover, the theory predicts the species abundance distribution across sites, which depends on the competitive balance of all species (Methods). To visualize these predictions, we keep the competitive balance of species 3 fixed and vary the ones of the first two (\figref{fig:balance-and-violins}B), and identify critical boundaries in competitive-balance space, partitioning it into distinct regions where only a given subset of species can survive.
\figref{fig:balance-and-violins}B shows that full coexistence is possible only within a limited region that encloses the equal balance condition.

A key insight is that equal competitive balance does not imply that species have identical traits. Rather, it reflects a global trade-off at the community level. Indeed, even in a simulation where species are assigned identical competitive balances but differ in their traits, they exhibit distinct abundance distributions and, in particular, mean abundances (\figref{fig:balance-and-violins}C). This demonstrates that competitive balance is not trivially linked to mean abundance, and it captures a fundamental ecological constraint instead.

In real metacommunities, a perfect match of competitive balances is neither expected to be possible nor is it required for coexistence. Our theory predicts that a group of competing species can coexist within the same metacommunity as long as their competitive balances are not too dissimilar.
Remarkably, this prediction aligns closely with observations from empirical data (\figref{fig:balance-and-violins}D-F). While the empirical effective dispersal strength and dispersal heterogeneity extracted from the VAA distributions have a large spread -- reflecting the high variability in the parameters describing effective dispersal strategies among species -- the situation is strikingly different for their competitive balances. Their values are tightly clustered around the mean, despite the broad range of individual species parameters. In other words, the competitive balances inferred for all species in the forest communities we analyze are closely aligned -- and it is this narrow distribution that enables their widespread coexistence.

\subsection*{Species vulnerability as a distance in competitive balance space}
\begin{figure}
		\centering
		\includegraphics[width=\textwidth]{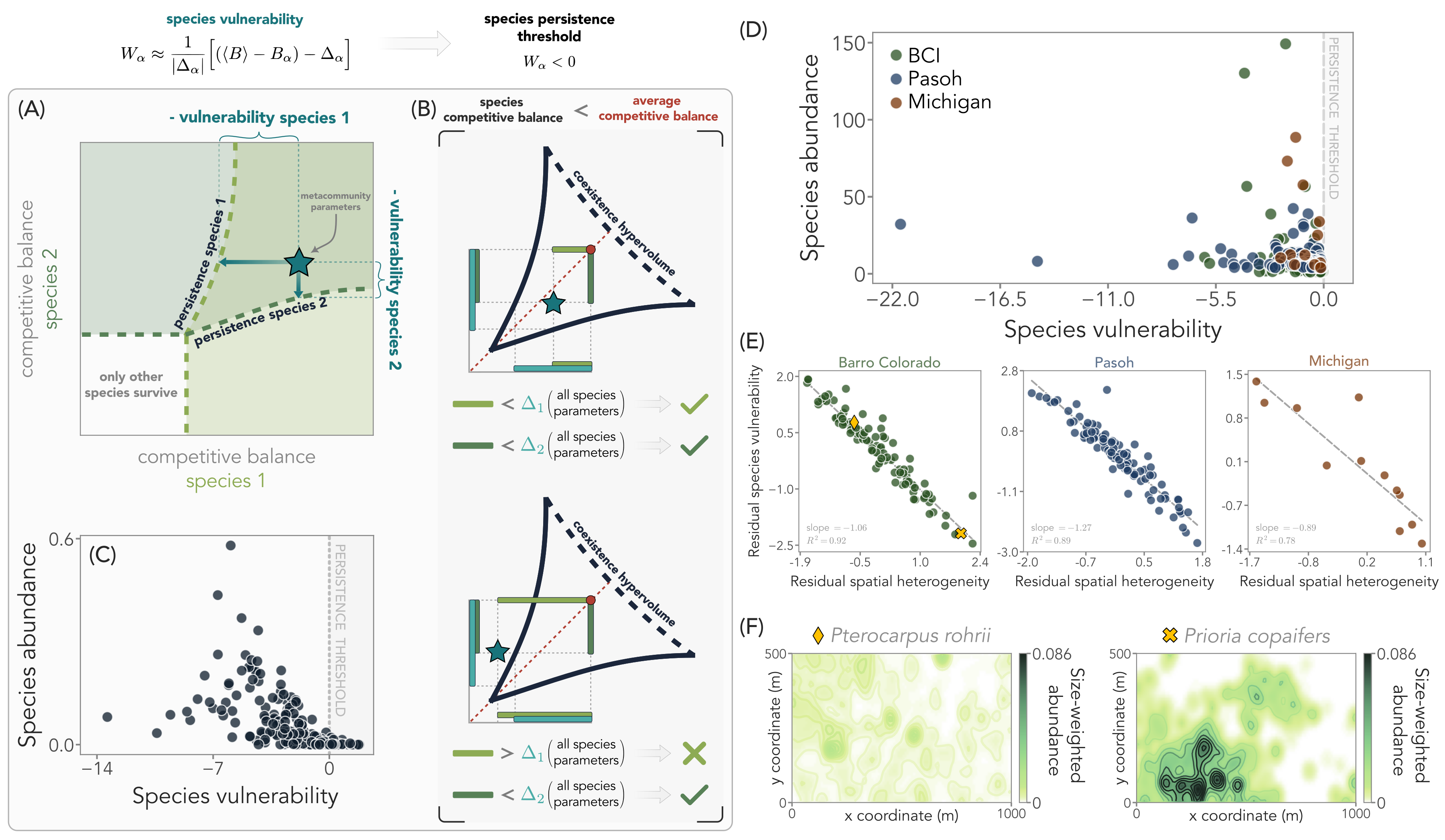}
		\caption{\footnotesize\textbf{Distance to persistence threshold within the competitive balance space defines vulnerability, which has no trivial relation to abundance and is strongly influenced by dispersal heterogeneity.}
		(\textbf{A})  The star marks the community's position in competitive balance space, here projected onto a two-species subspace. Dashed lines indicate species-specific critical boundaries, below which that species cannot persist in the community.  A species' (negative) distance from the boundary quantifies its vulnerability. (\textbf{B}) Schematic of practical vulnerability computation. The red circle marks the community average along the equal-balance line, serving as a baseline for perturbation analysis. The projection of the distance from the community's position to the red circle onto species 1's axis (green line) determines persistence: if this projection is smaller than $\Delta_\alpha$, the approximate distance from the circle to the critical boundary (cyan line), the species survives. In the top panel, both species persist in the larger metacommunity; in the bottom panel, species 1 is below the persistence threshold. (\textbf{C}):  Simulated metacommunity: stationary average abundance versus species vulnerability for randomly parameterized communities (Methods). (\textbf{D}): Scatter plot of species vulnerability versus abundance for the three forest datasets. All species exhibit negative vulnerability (i.e., persistence), and abundances span a broad range except near threshold. (\textbf{E}): Partial correlation (controlling for shared correlations with mean abundance, see Methods) between $\sigma_{p_\alpha}$, the spatial heterogeneity of the space-use distribution, and $W_\alpha$ of all forest datasets at $L=50$~m. The significant negative correlation between residuals indicates that, at fixed mean abundance, greater spatial variability reduces a species' vulnerability to competition. The two yellow markers indicate two representative species that display comparable total abundances but different spatial heterogeneity. (\textbf{F}) Spatial size-weighted distribution for the two representative species of the BCI forest. For \textit{Prioria copaifera}, the spatial heterogeneity is high ($\sigma_{p_\alpha}=0.038$), whereas for \textit{Pterocarpus rohrii} the spatial heterogeneity is lower ($\sigma_{p_\alpha}=0.003$). Their total stem-count is similar ($1348$ vs.~$1380$).}
		\label{fig:vulnerability}
\end{figure}

When the competitive balances of  species in a metacommunity diverge too far from each other, species with lower competitive balance may  not survive long-term competition. Our theory predicts that the \emph{vulnerability} of a species can be quantified in terms of how close it is to its competitive-exclusion threshold, which depends on the competitive balance of all other species. 
In principle, our theory allows to compute the extinction boundaries for each species as the competitive balance below which the species cannot survive in the community. Then, the distance from the metacommunity's position in the competitive balance space to its own persistence boundary provides a metric of how vulnerable a species is to competitive exclusion. We denote the vulnerability of species $\alpha$ with $W_{\alpha}$, defined such that $W_{\alpha}<0$ indicates that the species can persist in the metacommunity, while a positive $W_{\alpha}$ signals that it will eventually be outcompeted. This idea is illustrated in \figref{fig:vulnerability}A for two species.

In practice, however, the computation of vulnerability in this way becomes infeasible in species-rich communities, as the extinction boundaries depend not only on the exact values of all other species' competitive balances, but also iteratively on whether those species are above or below their own persistence thresholds in sub-communities  -- resulting in a combinatorial explosion of interdependent critical boundaries. To work around this difficulty, we developed an approximate measure which becomes increasingly accurate as the number of interacting species grows (Methods). We estimate the distance to the persistence boundary of a given species based on a perturbative expansion of the competitive balance using the equal-balance condition as a reference point. The equal-balance condition provides a baseline reference point from which the community competitive balance deviates. If the projection of this deviation onto the species' axis is greater than $\Delta_\alpha$, the distance between the baseline point and the critical boundary of species $\alpha$, the species cannot survive long-term competition.
The core idea is represented graphically in \figref{fig:vulnerability}B, where two possible cases are shown - one where both species can coexist with the community and one where one of them cannot survive. The explicit expression for the vulnerability of species $\alpha$ is given in the Methods and \tabref{tab:definitions}. The definition explicitly shows that
the vulnerability is not determined solely by a species' own traits, rather it depends on the composition of the entire community.

Numerical simulations of the model with randomly chosen species parameters show that species with positive vulnerabilities have a stationary abundance that approaches zero, consistent with the theory (\figref{fig:vulnerability}C). Interestingly, we observe that away from the persistence threshold, there is a substantial spread in the species abundances, suggesting that low abundance does not necessarily imply high vulnerability.

\subsection*{Less abundant does not always imply more vulnerable}
We can now apply our theory to calculate the vulnerability of species in the three forest plots. Consistent with our theoretical predictions, the results for all three datasets show that all species have vulnerability values below the persistence threshold --- $W_\alpha < 0$. This reinforces the conclusion that the observed communities lie within the coexistence region predicted by our theoretical approach and are stable  (\figref{fig:vulnerability}D). Remarkably, we find that abundance and vulnerability are strongly correlated only for species with vulnerability close to zero, i.e., close to the extinction threshold. Far from this threshold, instead, we find a large range for each species' average abundance, signaling that species with low average abundance are not necessarily at risk of competitive suppression, consistent with the observation made above that competitive balance depends on the entire community and it is thus not necessarily associated to low average abundance. These findings are consistent across all three datasets and are supported by numerical simulations of the model.

Away from the threshold, since average abundance is not a strong predictor of species vulnerability, we ask whether there are systematic relationships between a species' dispersal strategy, its vulnerability, and its mean abundance.
Our theoretical analysis (Methods) informs us that species with the same vulnerability and more evenly distributed dispersal kernels have higher average abundances, because the risk of oversaturation is locally reduced.
However, for two species with the comparable average abundance, this does not indicate what can explain differences in vulnerabilities. Although our theoretical framework does not allow for directly inverting vulnerability as an explicit function of dispersal heterogeneity, we can analyze this relationship in more detail. Specifically, we examine the partial correlation between $W_\alpha$, a species' vulnerability, $\sigma_{p_\alpha}$, the standard deviation of the spatial space-use fraction distribution, quantifying the heterogeneity of its spatial distribution. This partial correlation measures the association between these two variables after accounting for their shared dependence on mean abundance (Methods).
The results reveal a clear and significant negative correlation between $\sigma_{p_\alpha}$ and $W_\alpha$ in all three datasets (\figref{fig:vulnerability}E, detailed statistical analysis is reported in the SI sec. 8). This indicates that even if two species were to have the same average abundances but different spatial variability, the one with more heterogeneous dispersal would tend to be a superior competitor.
Intuitively, heterogeneous dispersal means that -- given a fixed colonization budget -- there will be habitat regions where a locally strong competitor is difficult to exclude, despite a reduction in overall abundance due to local saturation. This effect is visualized for two example species from the BCI dataset (\figref{fig:vulnerability}F) that have comparable total abundance ($\approx 2\%$ difference) but exhibit rather different spatial distributions and hence different vulnerabilities. This trade-off between total abundance, abundance variability, and vulnerability may represent a key mechanism underlying the coexistence of a large number of species.

\subsection*{Discussion}
We have validated a novel theoretical framework against three forest communities. By exploiting the exact solution of an underlying dynamical metacommunity model, we introduced a set of metrics that can be efficiently inferred from large-scale ecological data. While derived within a specific mathematical framework, these metrics admit model-independent interpretations, suggesting that their relevance may extend beyond the assumptions of the model. In particular, our work enables an intuitive interpretation of the relationships between vulnerability, abundance, and dispersal strategies within the metacommunity. These relationships represent non-trivial tradeoffs. On the one hand, at a given competitive balance species with a high abundance tend to have less spatially variable dispersal, allowing them to colonize a greater number of patches effectively, while species with more heterogeneous dispersal patterns (broader kernel distributions) will be more subject to crowding effects. On the other hand, at a given abundance, a species with a more spatially variable colonization strategy will be less vulnerable, because it can effectively create a spatial niche, from which it is harder to push out, buffering competitive suppression. These insights provide a mechanistic link between dispersal heterogeneity, abundance, and survival, illustrating how metacommunity structure emerges from individual-level dispersal strategies.

The idea that trait-derived quantities govern competitive outcomes has a long theoretical history, from resource-ratio theory \cite{Tilman1985} to modern trait-based coexistence frameworks \cite{MayfieldLevine2010,Bimler2018}. Our approach extends this lineage by inferring the relevant trade-off directly from spatial abundance data, without requiring prior trait measurements. In this sense, our measure of competitive balance can be viewed as a community-wide generalization of the competition-colonization trade-off \cite{levins1971regional,Levin1974dispersion,DurLev1998,mouquet2002coexistence,Miller2024allesina}. 
Although community-level manifestations of this trade-off have been reported \cite{levine2002coexistence,clark2004fecundity, spatial_coex_Hart_levine_2017}, their implications for coexistence in heterogeneous landscapes have remained unclear \cite{Vellend2017}. Here, we make this link explicit and show how such trade-offs shape species' risk of competitive exclusion. 
Frameworks that characterize coexistence as a feasibility domain in parameter space have enabled empirical applications, but rely on estimates of interaction coefficients or trait-based proxies \cite{Spaak2020}. In species-rich systems, such information is typically unavailable, and incomplete knowledge of niche dimensions can substantially affect the inference of interaction structure \cite{D_Andrea_2018}. By contrast, the analytical solution of our spatial model integrates dispersal and interaction effects at the community level, requiring only spatial census data for its application. Our vulnerability index therefore provides a flexible, context-dependent measure that accounts for the community composition. That is, a species with low vulnerability in one metacommunity may be at risk in another, depending on both species composition and spatial structure.

This context-dependent nature of our metrics could make our work suitable for several extensions. For instance, the introduction of an alien species with a specific competitive balance would lead to different coexistence or extinction scenarios of the entire community, depending on its relative position in the trade-off spectrum compared to the native community. This dynamic reconfiguration of species persistence underscores the interaction-dependent nature of a community's stability.
More generally, the framework could be extended to investigate hypothetical ecological disturbances such as the selective suppression of a single species, tracking how the vulnerability of the remaining species is altered and whether cascading exclusions analogous to a domino effect would emerge. These applications suggest how the framework may be used to probe the resilience and fragility of ecological networks under perturbations. The potential to extract information on temporal dynamics of biodiversity from spatial distributions of terrestrial species was already observed empirically \cite{Wilson_2004Kunin}. Our work provides a solid theoretical foundation for this type of inference.

Our results are not devoid of limitations. One is that our vulnerability metric does not provide information on the timescales over which different species approach their steady-state, but only predicts the final outcome. In our numerical simulations, we observed that these timescales are highly dependent on species-specific parameters, with some species stabilizing faster than others. This discrepancy suggests that transient dynamics, which are not captured by the vulnerability index, play a crucial role in shaping short- and medium-term ecological interactions. Understanding the factors that drive these variations in transient behavior is an important direction for further investigation \cite{comins1985dispersal,lavorel1994spatio}. 

Closely related to this, at very low abundances, demographic fluctuations are expected to become increasingly important. Recent work has highlighted the need to introduce probabilistic frameworks into coexistence theory \cite{Bowler2022}. In this context, important directions for future work include incorporating stochastic effects and accounting for temporal fluctuations \cite{Jops2023oDwyer,jops2025oDwyer} to move beyond deterministic proximity to exclusion, and exploring whether these can be quantitatively linked to the spatial buffering effect described here.

A further limitation is that distinct mechanisms of local competition and facilitation cannot be disentangled in our framework. At the level of description provided by our model, both contribute by amplifying or buffering the inferred kernel parameters. As recent works have highlighted the fundamental role of facilitation in driving coexistence \cite{McIntire2014,Losapio2021,Bimler2018,Bimler2025}, extending the framework to explicitly resolve facilitative interactions either through pairwise or trait-mediated effects \cite{Pichon2024,Buche2025} remains an important direction for future work. This extension is particularly relevant in systems where such interactions are expected to play a stronger or more structured role, and would likely require going beyond the current analytically tractable formulation of the model.

Overall, our study provides a tractable approach for studying metacommunity dynamics by bridging ecological theory and statistical physics, and provides a novel perspective on the fundamental principles governing biodiversity patterns and their resilience to perturbations.

\bibliography{DMFT-meets-forests}

\section*{Methods}

\subsection*{Model}
Our theory is based on a mathematical model describing the temporal dynamics of $S$ species inhabiting a landscape of $N$ sites, each with a finite amount of available space. The model tracks how the fraction of site $i=1\dots N$ occupied by species $\alpha$ changes with time. We denote this fraction as $p_{\alpha i}$. Being defined as a fraction, $0 \leq \sum_{\alpha=1}^S p_{\alpha i} \leq 1$ for all sites, which also implies  $0 \leq p_{\alpha i} \leq 1$ for all $i$ and $\alpha$.
The model dynamics are governed by the following set of differential equations:
\begin{equation}
	\der{p_{\alpha i}(t)}{t} = -e_{\alpha} p_{\alpha i}(t) + \left(1-\sum_{\beta=1}^{S}p_{\beta i}(t)\right)\sum_{j=1}^{N} K_{\alpha, ij} \, p_{\alpha j}(t).
	\label{eq:model-Grundgleichung-compact}
\end{equation}
Here, $K_{\alpha, ij}$ is a species-specific interaction-colonization kernel of species $\alpha$. Its diagonal elements capture the strength of local growth and competition processes, while the off-diagonal kernel elements encode the overall influence of all possible colonization processes through which patch $j$ can affect patch $i$, \ie the rate at which individuals of species $\alpha$ generated in patch $j$ can disperse through the environment and eventually colonize patch $i$. The term $(1-\sum_{\beta=1}^{S}p_{\beta i})$ represents the free space in patch $i$, enforcing saturation of available space.

The kernel matrix $\mathbf{K}_\alpha$ is an \emph{effective} species-specific interaction--colonization kernel. It quantifies both how species interact and how they disperse between patches. The kernel is not measured directly from field data, nor do we attempt to reconstruct each individual kernel entry $K_{\alpha,ij}$ separately. Doing so would require detailed knowledge of local demographic rates, dispersal between all pairs of patches, and interactions in every local environment, which is generally infeasible in species-rich and spatially heterogeneous systems. Instead, we assume that, for each species, the entries of $\mathbf{K}_\alpha$ are drawn from a statistical distribution $\mathcal{Q}^{(\alpha)}_{N}$ characterized by a small number of species-specific parameters. These parameters capture the variability in interactions and dispersal strategies among species. Although the kernel elements are random, we are able to infer the parameters of $\mathcal{Q}^{(\alpha)}_{N}$ indirectly from spatial abundance patterns for each species, by exploiting an explicit solution of the model.

At the level of mathematical structure, our equations align with classic metapopulation and metacommunity frameworks \cite{HanOva2000,mouquet2002coexistence}, and previous work has shown that such dispersal kernels can emerge as deterministic limits of stochastic models describing local birth, death, exploration, and colonization processes \cite{PadNic2024}. To gain insight on the possible ecological processes at play, one can rewrite \cref{eq:model-Grundgleichung-compact} in the following form:
\begin{align}
	\der{p_{\alpha i}(t)}{t} &=  K_{\alpha, ii} p_{\alpha i}(t) \Big(1-p_{\alpha i}(t)\Big) \label{eq:Grundgleichung-expanded-1} \\
	&- e_{\alpha i} p_{\alpha i}(t) \label{eq:Grundgleichung-expanded-2}  \\
	&-\sum_{\substack{\beta=1 \\ \beta\neq\alpha}}^{S}p_{\beta i}(t) K^{'}_{\alpha \beta, ii} \, p_{\alpha i}(t) \label{eq:Grundgleichung-expanded-3}\\
	&+ \left(1-\sum_{\beta=1}^{S}p_{\beta i}(t)\right)\sum_{\substack{j=1 \\ j \neq i}}^{N} K_{\alpha, ij}p_{\alpha j}(t).
	\label{eq:Grundgleichung-expanded-4}
\end{align}
Each term on the right side of the equation corresponds to a different process: local logistic growth with rate $K_{\alpha, ii}$ (eq.~\ref{eq:Grundgleichung-expanded-1}); linear death with rate $e_{\alpha i}$ (eq.~\ref{eq:Grundgleichung-expanded-2}); local effective competition term proportional to $K^{'}_{\alpha \beta, ii}$, that should be understood as the difference between competition and reduction of competition due to facilitation between species (eq.~\ref{eq:Grundgleichung-expanded-3}); and colonization, which depends on local space availability and dispersal from other sites with rate $K_{\alpha, ij}$ (eq.~\ref{eq:Grundgleichung-expanded-4}).
Under two simplifying assumptions, -- (i) the local death rate is patch-independent ($e_{\alpha i} \equiv e_{\alpha}$), and (ii) the local competition term is independent of the invading species and equals the local growth rate $K^{'}_{\alpha \beta, ii} \equiv K_{\alpha, ii}$ -- the equations reduce to \cref{eq:model-Grundgleichung-compact}.

We also note in passing that recasting our model in terms of a  spatial mean-field analysis would yield a mathematical structure that has some resemblance to that of classical hierarchical competition models \cite{tilman1994hierarchical}. In our model, however, we do not assume a competition hierarchy \textit{a priori}. Instead, coexistence trade-offs emerge from the interplay between model dynamics and quenched randomness. To uncover such emergent trade-offs, a mean-field analysis is not adequate. Instead, we resort to a dynamical mean-field approach, as explained below. 

\subsubsection*{Dynamical mean field theory}

In real ecosystems, estimating the colonization kernel directly is challenging. One simplification is to assume that in large, heterogeneous environments lacking macroscopic barriers, the kernel elements are randomly distributed, mirroring the underlying landscape heterogeneity. We adopt the assumption of ``homogeneous inhomogeneity," where all patches are statistically equivalent. Hence, we assume that the kernel elements of each species are realizations of a random variable $K^{(\alpha)}$ with probability distribution $\mathcal{Q}^{(\alpha)}_{N}$. Under this framework, we make the following scaling assumptions for the characteristic function of the kernel distribution in the limit of a large number of sites $N\to\infty$:
\begin{equation}
	\int \mathrm{d}K e^{-i K z} \mathcal{Q}^{(\alpha)}_{N}(K) = e^{F^{(\alpha)}(z) /N}.
\end{equation}
With this assumption, the full system reduces to a set of $S$ effective stochastic equations, one for each species \cite{AzaMar2024}:
\begin{equation}
	\der{{p}_{\alpha}(t)}{t} = -e_{\alpha} p_{\alpha}(t) + \left( 1 - \sum_{\beta=1}^{S} p_{\beta}(t) \right) \eta_{\alpha}(t),
	\label{eq:DMFT-main}
\end{equation}
where $\eta_\alpha(t)$ is a  stochastic process, the properties of which summarize the dynamics of the coupled system in one representative patch. This approach represents a non-Gaussian generalization of the usual dynamical mean-field theory (DMFT) \cite{AzaMar2024}. The generalized DMFT yields the following equations relating the statistical properties of $\eta_\alpha(t)$ to those of $p_\alpha(t)$  and the colonization kernel $K$:
\begin{equation}
	\left\langle \prod_{l=1}^{L} \eta_{\alpha}(t_{l}) \right\rangle_{C} =  {N} \left\langle {K}_{\alpha}^{L} \right\rangle_{C} \left\langle \prod_{l=1}^{L} p_{\alpha}(t_{l}) \right\rangle,
	\label{eq:SC-noise-main-def}
\end{equation}
where $\ev{.}_C$ indicates a $L$-th order cumulant and $\ev{.}$ on the right hand side indicates a moment of order $L$. Because the moments of $p_\alpha$ and the cumulants of $\eta_\alpha$ are interdependent, \eqnref{eq:SC-noise-main-def} needs to be solved self-consistently, which is in general a challenging problem.

\subsubsection*{Approximate stationary solution defines vacancy-adjusted abundance}

We focus on the long-term behavior of \eqnref{eq:DMFT-main}, and ask which species will persist. The stationary state $\left\{p^*_{\alpha}\right\}$ obeys the equation
\begin{equation}
	0 = - e_{\alpha} p_{\alpha}^{*} + \left( 1 - \sum_{\beta=1}^{S} p_{\beta}^{*} \right) \eta_{\alpha}^{*},
	\label{eq:p-star-main}
\end{equation}
where $\left\{\eta^*_{\alpha}\right\}$ is the stationary state of $\eta_\alpha(t)$. From now on, we will drop the asterisks to reduce notation clutter.
We define the  \emph{vacancy-adjusted abundance}   (VAA) discussed in the main text as $x_{\alpha} = \eta_{\alpha} / e_{\alpha}$
and solve \eqnref{eq:p-star-main} for it:
\begin{equation}
	x_{\alpha} = \frac{p_{\alpha}}{1 - \sum_{\beta} p_{\beta}},
	\label{eq:xstar-methods}
\end{equation}
which captures the ratio of abundance to the amount of free space available.
An important property of the VAA derived from the DMFT is that, in the limit of large-system size, its joint distribution is a product of individual distributions across different species \cite{AzaMar2024}, which greatly facilitates its comparison to empirical data.

Assuming that the cumulant generating function of the kernel  distribution, $F^{(\alpha)}(z)$, is analytic, the definition of $\eta_\alpha$, the self-consistent noise process in \eqref{eq:SC-noise-main-def} is equivalent to the following equation
\begin{equation}
	F_{\eta_\alpha}(z) = \Big\langle F^{(\alpha)} \big(z p_\alpha\big) \Big\rangle,
\end{equation}
where $F_{\eta_\alpha}(z)$ is the cumulant generating function of $\eta_\alpha$, and angular brackets indicate averaging over realizations of the noise process $\eta_\alpha$
To make analytical progress, we further assume that the statistical distribution of the kernel $\boldsymbol{K}_\alpha$ can be well approximated by a Gamma distribution with species-specific parameters $\delta_{\alpha}$ and $\beta_{\alpha}$, as discussed in the main text,
\begin{equation}
	K_{\alpha, ij} \sim \mathcal{Q}^{(\alpha)}_N(K) \equiv \frac{\beta_{\alpha}^{\frac{\delta_{\alpha}}{N}}}{\Gamma(\frac{\delta_{\alpha}}{N})} K^{-1 + \frac{\delta_{\alpha}}{N}} e^{-\beta_{\alpha} K} \Theta(K),
\end{equation}
where the shape parameters of the Gamma distributions, $\delta_{\alpha}$, are scaled with $N$ to have a consistent large-system-size limit. Then, assuming that the intensity of the self-consistent noise is not overwhelmingly strong, we can make the approximation $\ev{F^{(\alpha)}(z p_\alpha)} \approx F^{(\alpha)}(z \ev{p_\alpha})$, which leads to an explicit solution for the distribution of $x_\alpha=\eta_\alpha / e_\alpha$
\begin{equation}
	x_{\alpha, ij} \sim \mathcal{L}_{\alpha}(x) \equiv \frac{{\bar{\beta}}_{\alpha}^{{\delta_{\alpha}}}}{\Gamma({\delta_{\alpha}})} x^{-1 + {\delta_{\alpha}}} e^{-{\bar{\beta}}_{\alpha} x} \Theta(x),
	\label{eq:x-th-distr}
\end{equation}
where ${\bar{\beta}}_{\alpha}  = \beta_{\alpha} e_{\alpha} / \ev{p_{\alpha}}$, and $\ev{p_{\alpha}}$ denotes the abundance of species $\alpha$ averaged across sites. Consequently, the VAA itself follows a Gamma distribution, with parameters distinct from -- but related to -- those of the colonization kernel via the above relation.  We can now explicitly relate $\bar{\beta}_{\alpha}$ to the kernel parameters by starting from the following  equation for $\ev{p_\alpha}$, obtained by inverting \eqref{eq:xstar-methods} and  applying the definition of the spatial average:
\begin{equation}
	\ev{p_{\alpha}} = \int \prod\limits_{\mu=1}^{S} \left[ d x_{\mu} \, \mathcal{L}^{(\mu)} (x_{\mu}) \right] \frac{{x_{\alpha}}}{1 + \sum_{\gamma} {x_{\gamma}}}.
	\label{eq:self-consistent-mean}
\end{equation}
Manipulating this last equation yields a system of integral equations  that links the set of all $\{\bar{\beta}_{\alpha}\}_{\alpha=1 \dots S}$  to the inverse of the \emph{competitive balance} of species $\alpha$:
\begin{equation}
	B_{\alpha}^{-1} = \int_{0}^{\infty} d\lambda \, \frac{e^{-\lambda
			- \sum_{\gamma} \delta_{\gamma} \ln \left(1 + {\lambda}/({{\bar{\beta}}_{\gamma} )}\right)}}{ \left.1 + {\lambda}/({{\bar{\beta}}_{\alpha} })\right.},
	\label{eq:B-alpha-main}
\end{equation}
where the competitive balance itself is defined as
\begin{equation}
	B_\alpha \equiv   \frac{\delta_{\alpha} }{\beta_{\alpha} e_{\alpha}}.
	\label{eq:CB-definition}
\end{equation}
We now consider the case in which only one species is present. We set in \eqref{eq:self-consistent-mean} the abundances of all species except $\alpha$ to zero, and analyze the integral on the right-hand side in the limit $\ev{p_\alpha} \to 0^+$. We posit that the transition between persistence and extinction is smooth. Through this approach, we find that the transition occurs - that is $\ev{p_\alpha} = 0^+$ when the ratio $\beta_\alpha e_\alpha / \delta_\alpha$ equals unity. In other words, the species can survive in isolation when $B_\alpha > 1$.

\subsubsection*{(Co)existence transition boundaries}

As mentioned above, the persistence condition $B_{\alpha}>1$ holds for a species in isolation. When other species are present, this critical transition to persistence depends on the competitive balance of all species. Specifically, the critical values of $\{ B_{{\rm crit},\alpha\varphi} \}_{\phi=1\dots  S}$ marking the transition of species $\alpha$ from local extinction to survival when all other $S-1$ are present form a $S-1$ dimensional manifold described by the following parametric equations
\begin{align}
	B_{{\rm crit},\alpha\varphi}^{-1} &= \int_{0}^{\infty} d\lambda \, e^{-\lambda} \prod_{\gamma \neq \alpha}^{S} \left( 1 + t_{\gamma} \lambda \right)^{-\epsilon_{\gamma\varphi}}
	\label{eq:critical-manifolds-1}
	\\
	\epsilon_{\gamma\varphi} &=
	\label{eq:critical-manifolds-2}
	\begin{cases}
		\delta_{\varphi} + 1, & \gamma = \varphi \\
		\delta_{\gamma}, & \gamma \neq \varphi
	\end{cases}
\end{align}
where $0 < t_{\gamma}< \infty$ are the parametric descriptors of the manifold and $\gamma = 1, 2 \dots \alpha - 1, \alpha +1, \dots, S$. These critical boundaries are obtained from \eqnref{eq:B-alpha-main} evaluating the limit $1 / {\bar{\beta}_{\alpha}} \to 0^{+}$, which corresponds to $\ev{p_\alpha} \to 0$. Boundaries in \figref{fig:balance-and-violins}A and B were plotted using \eqnref{eq:critical-manifolds-1} and (\ref{eq:critical-manifolds-2}) in the case $S=3$, with $B_3=2$, $\delta_1=1.5$, $\delta_2=3$, $\delta_3=2$.

If one or more species are extinct, then they need to be excluded from the product term. If one species goes extinct, there are $S$ new possible transition manifolds (each of dimension $S - 1$). Similarly, if two species are extinct, one obtains a further transition manifold for each of the $\binom{S}{2} = \frac{S(S-1)}{2}$ possible pairs of extinct species. This pattern continues, leading to a combinatorial explosion of possible transitions to track.
We note that, in principle, recomputing these transition manifolds iteratively could be used to predict how the removal or introduction of species affects community stability, as discussed in the outlook (Discussion). However, the number of such boundaries grows combinatorially with the number of species $S$, rendering this approach impractical in species-rich ecosystems and motivating the use of the vulnerability index as a scalable alternative.

\subsubsection*{Perturbative expansion leads to vulnerability definition}

To circumvent the problem of the combinatorial explosion of transition boundaries, we start from the point in the competitive balance space defined by the condition $B_{\alpha} = B_{00}$, \ie the scenario in which the competitive balance of all species are equal to the same value.
Then, we consider the case of a large number of species $S \gg 1$. We define $Z_{\alpha}\equiv 1 /B_{\alpha}$ for convenience, and the rescaled abundance $y_\alpha = \ev{p_\alpha} S$, which stays finite in the limit $S\to\infty$ as opposed to $\ev{p_\alpha}$, which goes to zero because of the condition $\sum_{\beta=1}^Sp_\beta \leq 1$.
We now make the following expansion in powers of $1/S$
\begin{equation}
	Z_{\alpha} = Z_{\alpha 0} \left(1 - \frac{Z_{\alpha1}}{S} + \dots \right),
	\label{eq:Zalpha-expansion}
\end{equation}
where $Z_{\alpha 0}$ has been allowed to depend on $\alpha$ for generality, but will end up being the equal-balance condition discussed in the main text. Given this ansatz, we also expect the rescaled average abundances to scale as
\begin{equation}
	y_\alpha = y_{\alpha 0} \left(1 + \frac{y_{\alpha 0}}{S} + \dots \right).
\end{equation}
Then, we substitute \eqnref{eq:Zalpha-expansion} into the left hand side of \eqnref{eq:B-alpha-main}, and $\ev{p_\alpha} = y_\alpha / S$ into its right hand side, which we then expand in powers of $1/S$. Matching leading order terms yields the following consistency condition for $Z_{\alpha 0}$
\begin{equation}
	Z_{\alpha 0} =  Z_{00} > 1,
\end{equation}
and the following expression for the zero order of $y_{\alpha}$:
\begin{equation}
	y_{\alpha 0} = \frac{\beta_{\alpha} e_\alpha (Z_{\alpha 1} - \Delta )}{Z_{00}}; \quad
	\Delta = \frac{ \left[ {\beta}_{\alpha}e_{\alpha} Z_{\alpha1} \right]_{\alpha} - Z_{00} \left( 1 - Z_{00} \right)}{\left[ {\beta}_\alpha e_{\alpha} \right]_{\alpha}},
\end{equation}
where $\left[ \cdot \right]_\alpha$ indicates an average over species. The persistence condition for species $\alpha$ is $y_{\alpha 0}>0$, which, using \eqnref{eq:Zalpha-expansion}, leads to
\begin{equation}
	y_{\alpha 0}>0 \, \Leftrightarrow \, \frac{1}{|\Delta|}\biggl(\Delta -  S \frac{\left[Z_\alpha\right]_\alpha- Z_{\alpha} }{\left[Z_\alpha\right]_\alpha} \biggr) < 0.
	\label{eq:species-vulnerability-0}
\end{equation}
In \eqnref{eq:species-vulnerability-0}  we identify $Z_{00} = [ Z_\alpha ]_{\alpha}$ as the baseline point for the expansion, as it minimizes the distances from it and is most consistent with the ansatz the expansion is based on. The inequality in \eqnref{eq:species-vulnerability-0} corresponds to the definition of the vulnerability $W_\alpha$.

To remain consistent with species-specific parameters, we rewrite the vulnerability in terms of the species' competitive balances $B_\alpha = Z^{-1}_\alpha$ to obtain the intuitive form of the vulnerability discussed in the main text. To this end, we make the approximation $[1 / B_\alpha]_\alpha \approx 1 / [B_\alpha]_\alpha$, which holds when $B_\alpha$ are close to each other, as required for coexistence when $S$ is large. In this way, we can rearrange \eqnref{eq:species-vulnerability-0} obtaining
\begin{equation}
	W_\alpha \approx \frac{1}{|\Delta_\alpha|}\biggl(\big(\left[B_\alpha\right]_\alpha- B_{\alpha}\big)  -\Delta_\alpha \biggr) < 0,
	\label{eq:species-vulnerability-1}
\end{equation}
where $B_\alpha = \delta_\alpha/\beta_\alpha e_\alpha $ and $\Delta_\alpha = -\Delta \cdot B_\alpha / S$, thereby connecting vulnerability and species-specific dispersal parameters.

\subsection*{Numerical simulations}

Numerical simulations of the model were carried out by integrating Eq.~\eqref{eq:model-Grundgleichung-compact} using an adaptive Runge-Kutta scheme implemented in custom C++ code based on GSL scientific libraries. Simulations were run until the system reached a stationary state, which was assessed by checking that the maximum change (across species and sites) in local abundance $| p_{\alpha i}(t + 1) - p_{\alpha i}(t)|$ fell below a precision threshold $\theta=10^{-5}$.

The simulation results shown in \figref{fig:balance-and-violins}A (inset) were obtained by simulating a metacommunity of five species with equal $B_\alpha$ but with $\delta_{\alpha}$ values evenly spaced in the interval $(2, 8)$, progressing to stationarity in a landscape of $N=2000$ sites. Only three of the five species are shown for visualization clarity. Results were averaged over 1000 realizations of the interaction-dispersal kernel $\mathbf{K}$.
The results shown in \figref{fig:balance-and-violins}B are based on simulations of three species with $\delta_1 = 1.5$, $\delta_2 = 3$, $\delta_3 = 2$ and $B_3 = 2$, over $N=800$ sites. The competitive balance between the first two species was systematically varied by changing $\beta_1$ and $\beta_2$. For each grid point, results were averaged over ten realizations of $\mathbf{K}$.
Finally, the results shown in \figref{fig:vulnerability}C were obtained from 40 realizations of a community of $S=6$ species distributed over $N=1200$ sites, with each species assigned randomly chosen values of $B_\alpha$ and $\beta_{\alpha}$. The values of $\beta_{\alpha}$ were uniformly distributed in the interval $(0.5, 3.0)$, while $B_\alpha$ values were randomly drawn from intervals centered around either $B_{00} = 1.25$ or $B_{00} = 1.66$, with varying spreads in the range $(0.5, 1.0) / S$ to mimic variability in natural communities.

\subsection*{Forest data analysis}

We partition each forest plot into square subplots of side length \(L\). For the results presented in the main text, we use \(L = 50\)~m, selected from a range of spatial scales (10--125~m for BCI and Pasoh, and 10--100~m for Michigan, the latter limited by the need to tile its non-rectangular geometry).

For a given spatial scale $L$, we estimate the effective space within patch $i$ occupied by species $\alpha$, denoted $a_{\alpha i}$, using two alternative metrics. The simpler \emph{stem-count} metric sets $a_{\alpha i} = n_{\alpha i}$, the number of individuals, implicitly assuming equal space occupancy. The \emph{DBH-weighted} metric instead defines
$a_{\alpha i} = \sum_j d_{\alpha i,j}^{4/3},$ where $d_{\alpha i,j}$ is the diameter at breast height (DBH, in cm) of the $j$-th individual of species $\alpha$ in subplot $i$. The exponent $4/3$ follows from the allometric scaling between trunk diameter and projected crown area, which is well established both theoretically and empirically \cite{Simini2010pnas,MartinezCano2019}, so that $a_{\alpha i}$ approximates the effective canopy space occupied by species \(\alpha\).
We verified that both metrics yield consistent estimates of the effective per-patch carrying capacity, and that converting the DBH-weighted carrying capacity to actual canopy area using allometric scaling coefficients for BCI \cite{MartinezCano2019} yields an estimated number of canopy layers consistent with those reported for humid forests \cite{Terborgh1985,DAndrea2020} (see SI, sec.~1). Based on a quantitative goodness-of-fit comparison across spatial scales and metrics (SI, Sec.~1.1), we adopt the DBH-weighted metric as the primary abundance measure and $L = 50$~m as the standard spatial scale for all main-text analyses. For model comparison, however, we use the stem-count metric. Incorporating DBH-weighting in the null models would require additional assumptions on the sampling of individual sizes and introduce further arbitrariness, while also increasing the number of parameters. Using stem counts therefore ensures a consistent and fair comparison across models.

Species selection for the analysis was based on two criteria, balancing sufficient abundance and adequate spatial sampling. Starting from all species with at least one recorded alive stem, we retained the most abundant species until their cumulative stem-count abundance reached at least $95\%$ of the total abundance in the plot. We then excluded species present in fewer than $10\%$ of subplots, as their VAA distributions cannot be reliably estimated due to insufficient spatial sampling. This procedure yielded 101 (BCI), 396 (Pasoh), and 12 (Michigan) species at $L = 50$~m.

We define the per-patch space use as $p_{\alpha i} = a_{\alpha i}/M$, where $M$ is a free parameter representing the total carrying capacity of each subplot, assumed to be spatially homogeneous (we investigated the robustness of this assumption in the SI, sec.4). To determine $M$, we scan a coarse grid of candidate values, starting from $M_0 = \max_i \bigl\{\sum_\beta a_{\beta i}\bigr\} + 1$, the minimum value ensuring a positive denominator in the VAA, and incrementing upward. We then refine the search around the best coarse candidate using a finer grid.
For each candidate $M$, we compute the VAA for all species and patches using
\begin{equation}
	x_{\alpha i} = \frac{p_{\alpha i}}{1 - \sum_{\beta} p_{\beta i}},
\end{equation}
and fit, independently for each species, the empirical cumulative distribution of non-zero VAA values to the theoretical Gamma distribution, \cref{eq:x-th-distr}, via least squares. The optimal $M$ is selected by minimizing the total CDF mean-squared error across all selected species. This procedure is applied for both the stem-count and DBH-weighted metrics.

\subsubsection*{Species-specific dispersal parameters}
After determining the optimal value of $M$, we fitted the VAA distributions of each species at all levels of coarse-graining using \eqnref{eq:x-th-distr}. This allowed us to obtain the values of $\delta_{\alpha}$ and $\bar{\beta}_{\alpha}$. A detailed GOF analysis for all $>500$ species at all spatial scales is presented in the SI, together with visualization for 212 species (100 + 100 + 12 from the three datasets).  In the main text and in \figref{fig:balance-and-violins}, we refer to $\delta_{\alpha}^{-1}$ as \emph{dispersal heterogeneity}  because the shape parameter of the kernel distribution determines the heterogeneity of the dispersal matrix -- the coefficient of variation of the Gamma distribution is $\sqrt{\delta^{-1}}$. In the same figure and context, we refer to $\bar{\beta}_{\alpha}^{-1}$ as \emph{effective dispersal strength} reflecting the relation  $\bar{\beta}_{\alpha}^{-1}  =  \ev{p_{\alpha}} \cdot \beta_{\alpha}^{-1}$, which shows that $\bar{\beta}_{\alpha}^{-1}$ is proportional to the scale of the interaction-dispersal kernel distribution, multiplied by the species average abundance, thus motivating its name.
Finally, we employed \eqnref{eq:B-alpha-main} to compute each species' competitive balance, from which the value of $\beta_{\alpha}$ was obtained. Predictive accuracy was assessed by out-of-sample cross-validation, as described in the next subsection. We used \eqnref{eq:species-vulnerability-1} to compute the species vulnerability shown in \figref{fig:vulnerability}C,D.

\subsubsection*{Out-of-sample predictive accuracy}

To assess whether the theoretical prediction for the mean patch occupancy $\langle p_\alpha \rangle$ generalizes to unseen data, we performed a spatial cross-validation across all three datasets. For each of 20 independent replicates, we randomly split the subplots of each plot into two equal halves. We fitted the model on the training half, obtaining the species-specific parameters $\delta_\alpha^{\mathrm{tr}}$, $\bar{\beta}_\alpha^{\mathrm{tr}}$, and the global carrying capacity $M^{\mathrm{tr}}$, as described above. Using these parameters, we computed the theoretical mean patch occupancy as
$\langle p_\alpha \rangle^{\mathrm{th}} = ({\delta_\alpha^{\mathrm{tr}}\, B_\alpha}/{\bar{\beta}_\alpha^{\mathrm{tr}}})\,(1 - \pi_{0,\alpha}),$
where \(B_\alpha\) is the competitive balance integral \cref{eq:B-alpha-main}, evaluated using the training-half parameters for all selected species, and $(1 - \pi_{0,\alpha})$ is the fraction of non-zero patches, accounting for the absence of probability mass at zero in the Gamma distribution.
Results are summarized by taking the median across the 20 splits. The comparison is shown in \figref{fig:VAAD-fits}B. Additional out-of-sample validation analyses of the inferred parameters are reported in the SI (Sec.~2).

\subsubsection*{Model comparison with Negative Binomial nulls}
We benchmarked the our framework against a family of Negative Binomial (NegBin) null models that reproduce overdispersed abundance fluctuations but assume species independence and no spatial coupling \cite{Volkov2003,Linden2011,tovo2017upscaling-125}. We considered three variants of increasing parsimony: (i) a species-specific Negative NegBin, with independent mean and dispersion parameters $(\mu_\alpha, r_\alpha)$ for each species; (ii) an intermediate model with species-specific means $\mu_\alpha$ and a shared dispersion parameter $r$; and (iii) a fully neutral model with both $\mu$ and $r$ shared across species. Including the effective community capacity parameter, these models comprise $2S+1$, $S+2$, and $3$ parameters, respectively, compared to $2S+1$ parameters in our DMFT framework. Parameters were estimated by maximum likelihood (see SI, Sec.~6).

To ensure a consistent comparison, all models were evaluated in a common rescaled abundance space $x_s$. For each fitted model, we generated synthetic datasets matching the empirical dimensions and computed the Kolmogorov--Smirnov (KS) statistic between the observed and simulated $x_s$ distributions (restricted to positive values). Predictions from our model were sampled directly in $x_s$, whereas NegBin samples were generated in count space and transformed using the same rescaling as the data. KS statistics were obtained over repeated synthetic realizations and summarized across species, providing a relative measure of goodness of fit that enables direct comparison across model classes.
Because the rescaling depends on an effective community capacity, this parameter was estimated independently for each model to avoid circularity. In particular, the NegBin models were assigned their own capacity parameter, inferred by minimizing discrepancies between empirical and model cumulative distributions, ensuring that each model was evaluated under its optimal scaling.

In addition to this relative comparison, we assessed absolute goodness of fit using a Cram\'er--von Mises--type statistic computed directly from empirical distributions. For each species, we measured the mean squared difference between the empirical CDF of positive abundances and the fitted model CDF, scaled by the number of observations; under a well-specified continuous model this quantity has an expected value of $1/6$, providing a reference scale for interpretation. This metric does not rely on synthetic sampling and complements the KS-based comparison. All analyses were performed on stem-count data to ensure consistency with the discrete NegBin formulation. Further details are provided in the SI.

\subsubsection*{Functional guild analysis of VAA parameters}
To assess whether the VAA parameters reflect differences in life-history strategies, we compared species-specific parameter estimates across functional guilds in the BCI dataset. Guild assignments (shade-tolerant, gap-exploiting, and intermediate) were taken from ref.~\cite{Comita2007}.
Statistical comparisons were performed using two-sided Mann--Whitney U tests between guilds. To increase statistical power, we focused on an aggregated comparison between gap-exploiting and intermediate species versus shade-tolerant species, while also reporting the full three-guild analysis in the SI. P-values were corrected for multiple comparisons using the Holm step-down procedure.
Parameter distributions were summarized using medians and interquartile ranges, and all analyses were performed at the species level. Further details on sample sizes, full pairwise comparisons, and robustness of the results are provided in the SI.

\subsubsection*{Statistical analysis of vulnerability vs.~spatial heterogeneity}
In the main text, we present the partial correlation between $\sigma_{p_\alpha}$ and $W_\alpha$, while controlling for the effect of the mean abundance of each species. To this, we first performed a linear regression of the log-transformed $(-W_\alpha)$ against $\ev{n_\alpha}$, and another of $\ln \sigma_{p_\alpha}$ against $\ln \ev{n_\alpha}$. We then regressed the residuals of these two models, denoted $e_{W_\alpha}$ and $e_{\sigma_\alpha}$, respectively, to quantify the residual correlation between $W_\alpha$ and $\sigma_{p_\alpha}$ not explained by the control variable (the mean abundance). All regressions were conducted using the open-source Python library \texttt{statsmodels}, relevant code is provided in the code capsule~\cite{codeocean_capsule}. Figures and full details on this statistical analysis, for all three datasets and the model, are provided in the SI, sec.~8

\subsection*{Software}
All analyses were performed in Python~3.12, using \texttt{numpy}~1.26.4 for array operations, \texttt{scipy}~1.13.1 for numerical integration, optimisation and statistical distributions, \texttt{pandas}~2.2.2 for data handling, \texttt{statsmodels}~0.14.2 for the regression and partial-correlation analyses, and \texttt{matplotlib}~3.8.4 for figure generation. Version-pinned requirements and the complete analysis pipeline are provided in the Code Ocean capsule~\cite{codeocean_capsule}.

\subsection*{Code availability}
All code required to reproduce the analyses and figures in this paper is provided in the Code Ocean capsule~\cite{codeocean_capsule}. A user-friendly interface to the parameter extraction and vulnerability metric computation, implementing the core calculation described here, is additionally available as the open-source \texttt{vulntool} package at \url{https://github.com/davideb321/vulntool}.

\backmatter

\bmhead{Acknowledgements}

We thank the photographers and institutions who contributed images via iNaturalist and GBIF under Creative Commons or public-domain (CC0) terms. Full image credits and license details are provided in Table~\ref{tab:photo_credits}.
DB, SA, AM were supported by the Italian Ministry of University and Research (project funded under the National Recovery and Resilience Plan (NRRP), Mission 4, Component 2 Investment 1.4 - Call for tender No. 3138 of 16 December 2021, rectified by Decree n.3175 of 18 December 2021 of Italian Ministry of University and Research funded by the European Union -- NextGenerationEU; Award Number: Project code CN\_00000033, Concession Decree No. 1034 of 17 June 2022 adopted by the Italian Ministry of University and Research, CUP C93C22002810006, Project title ``National Biodiversity Future Center - NBFC"). SS has been funded by the European Union’s Horizon 2020 research and innovation program, project BIOcean5D with grant agreement No. 101059915. PP was supported by the Swiss National Science Foundation (SNSF) through the Swiss National Centre of Competence in Research (NCCR) Microbiomes (grant number 51NF40\_225148). The work in this paper was supported by a gift from William H. Miller III.
  \begin{table}[t]
	\centering
	\caption{Photo credits for Fig.~2. Species photographs are reproduced from iNaturalist and GBIF under the licenses listed. CC BY 4.0: \url{https://creativecommons.org/licenses/by/4.0/}. CC0 images are released under a public-domain dedication (no rights reserved).}
	\label{tab:photo_credits}
	\footnotesize
	\begin{tabular}{|l|l|l|l|}
		\hline
		Species & link & user & License \\
		\hline
		\textit{Faramea occidentalis} & \url{https://www.inaturalist.org/photos/218529068} & Ana Nuño & CC0 \\
		\textit{Garcinia intermedia}  & \url{https://www.inaturalist.org/photos/16325072} & Alexis López Hernández & CC BY 4.0 \\
		\textit{Protium panamense} & \url{https://www.inaturalist.org/photos/56233276} & Letizia Weichgrebe & CC BY 4.0 \\
		\textit{Xerospermum noronhianum} & \url{https://www.gbif.org/occurrence/4897553870} & Royal Botanic Gardens, Kew & CC BY 4.0 \\
		\textit{Ardisia crassa} & \url{https://www.gbif.org/occurrence/5130154644} & Royal Botanic Gardens, Kew & CC BY 4.0 \\
		\textit{Xanthophyllum eurhynchum} & \url{https://www.gbif.org/occurrence/5060737580} & Royal Botanic Gardens, Kew & CC BY 4.0 \\
		\textit{Prunus serotina} & \url{https://www.inaturalist.org/photos/631844036} & Tommy & CC0 \\
		\textit{Acer rubrum} & \url{https://www.inaturalist.org/photos/631489904} & Stephen Lewis & CC0 \\
		\textit{Amelanchier arborea} & \url{https://www.inaturalist.org/photos/502104628} & Marc-Aurèle Vallée & CC0 \\
		\hline
	\end{tabular}
	\vspace{0.5cm}
 \end{table}

\bmhead{Supplemental Information}
Contact email for inquiries regarding the Supplemental Information: davide.bernardi@unipd.it

\end{document}